\documentclass{tran-l}

\theoremstyle{definition}

\theoremstyle{remark}

\numberwithin{equation}{section}

\begin{document}

\title[Energy Distribution of a  Stringy Charged Black Hole]
{Energy Distribution of a Stringy Charged Black Hole}

\author{ Ragab M. Gad}
\address{ Mathematics Department, Faculty of
Science, Minia University, El-Minia, Egypt}

\email{ragab2gad@hotmail.com}






\dedicatory{}


\Large{
\begin{abstract}

The energy distribution associated with a stringy
charged black hole is studied using M{\o}ller's
energy-momentum complex. Our result is reasonable
and it differs from that known in literature
using Einstein's energy-momentum complex.
\end{abstract}

\maketitle
\section{Introduction}

Ever since the Einstein's energy-momentum complex, used for
calculating energy and momentum in a general relativistic system,
many attempts have been made to evaluate the energy distribution
for a given space-time, for instance, Tolman \cite{T34}, Landau
and Lifishitz \cite{LL87}, Papapetrou \cite{P48} and Weinberg
\cite{W72}. These definitions only give meaningful results if
the calculations are performed in "Cartesian" coordinates.
M{\o}ller \cite{M58} constructed an expression which enables one to
evaluate energy in any coordinate system.
\par
The main problem of these prescriptions is that whether or not
they give the same energy distribution for a given space-time.
Virbhadra \cite{V99} investigated the most general non-static
spherically symmetric space-times, using the Einstein,
Landau and Lifshitz, Papapetrou, and Weinberg
prescriptions, and he found  that the definitions of energy distribution
disagree in general. Recently, Xulu \cite{grO0} used the M{\o}ller
energy momentum expression to compute the energy distribution in
these space-times and compared his result with one obtained by
Virbhadra \cite{V99} in Einstein,  Landau and Lifshitz, Papapetrou,
and Weinberg prescriptions. He found that the
definitions of Einstein and M{\o}ller disagree in general.
\par
In this paper we use the M{\o}ller energy momentum expression to
compute the energy distribution associated with a stringy charged
black hole. The energy distribution of a stringy charged black
hole studied first by Virbhadra and Parikh \cite{VP93} in Einstein's
prescription
and followed by Xulu \cite{X98}, using
Tolman's energy-momentum complex. They obtained the same value of energy
distribution.
It is worth investigating whether or not other
definition of energy give the same result as obtained by them.
\par
Through this paper we use $G = 1$ and $c = 1$ units and follow the
convention that Latin indices take value from 0 to 3 and Greek
indices take value from 1 to 3.

\section{Energy in M{\o}ller's Prescription}
It is interesting to evaluate the energy distribution of a static
spherically symmetric charged black hole. The line element
represented this space-time is given by \cite{GHS91}
\begin{equation} \label{2.1}
ds^2 = (1 - \frac{2M}{r})dt^2 - (1- \frac{2M}{r})^{-1}dr^2 - (1 -
\frac{\alpha}{r})r^2(d\theta^2 + \sin^{2}\theta d\phi^2),
\end{equation}
where $$ \alpha = Q^2\frac{\exp (-2\Phi_{0})}{M}, $$ $M$ and $Q$ are,
respectively, mass and charged parameters; $\Phi_{0}$ is the
asymptotic value of dilation field.\\
The metric (\ref{2.1}) is almost identical to the Schwarzschild metric.
The only difference is that the areas of the spheres of constant $r$
and $t$ depend on the charge $Q$ \cite{GHS91}.\\
The determinant and the
non-zero contravariant components of the line element (\ref{2.1})
are
\begin{equation}\label{2.2}
\begin{split}
g & = -(r - \alpha)^2r^2\sin^2\theta,\\ g^{00} & = \frac{r}{r -
2M}, \\ g^{11} & = - \frac{r - 2M}{r},\\ g^{22} & = - \frac{1}{r(r
- \alpha )},\\ g^{33} & = - \frac{1}{r(r - \alpha)\sin^2\theta}.
\end{split}
\end{equation}

The M{\o}ller's energy-momentum complex $\Theta^k_i$ is given by
\begin{equation}\label{2.3}
\Theta^k_i = \frac{1}{8\pi}\chi ^{kl}_{i,l},
\end{equation}
where the antisymmetric superpotential $\chi^{kl}_i$ is
\begin{equation}\label{2.4}
\chi^{kl}_i = - \chi^{lk}_i = \sqrt{-g}\big( \frac{\partial
g_{in}}{\partial x^m} - \frac{\partial g_{im}}{\partial x^n}\big)
g^{km}g^{nl},
\end{equation}
$\Theta^0_0$ is the energy density and $\Theta^0_{\alpha}$ are the
momentum density components. Also, $\Theta^k_i$ satisfies the
local conservation laws:
\begin{equation}\label{2.5}
\frac{\partial\Theta^k_i}{\partial x^k} = 0
\end{equation}
The energy component $E$ is
\begin{equation}\label{2.6}
\begin{split}
E & = \int\int\int{\Theta^0_0 dx^1 dx^2 dx^3}\\ & =
\frac{1}{8\pi}\int\int\int{\frac{\partial \chi_{0}^{0l}}{\partial
x^l} dx^1 dx^2 dx^3}.
\end{split}
\end{equation}
For the line element (\ref{2.1}) the $\chi_{0}^{01}$ component is
given by
\begin{equation}\label{2.7}
\chi_{0}^{01} = \frac{2M}{r}(r - \alpha )\sin\theta,
\end{equation}
which is the only required component of $\chi_{0}^{0l}$.\\
From equations (\ref{2.6}) and (\ref{2.7}) and applying the Gauss
theorem, we obtain the energy distribution
\begin{equation}\label{2.8}
E = \frac{M}{r}(r -\alpha).
\end{equation}
\section{Discussion}
Several examples of charged black holes have been investigated and
their energy distribution has been obtained. The value of obtained energy
distribution
depends on the mass $M$ and the charge $Q$, for instance, in
Kerr-Newman $E_{Tol} = E_{LL} = \frac{Q^2}{R}( \frac{a^2}{3R^2} +
\frac{1}{2})$ \cite{V90} and for the Reissner-Nordstr\"{o}m,
several definitions of energy give $E =  M - \frac{Q^2}{2r}$ (see in
\cite{T83, H94, ACV96}), whereas $E_{M{\o}l} = M - \frac{Q^2}{r}$ 
\cite{V90a}.\\
Chamorro and Virbhadra \cite{CV96} and Xulu \cite{Xgr-qc/00} showed,
using Einstein and M{\o}ller prescriptions, that the energy distribution of
charged dilation black hole \cite{GHS91} depends on the value $\lambda$;
$E_{Einst} = M - \frac{Q^2}{2r}(1 - \lambda^2)$,
$E_{M{\o}l} = M - \frac{Q^2}{r}(1 - \lambda^2)$, where a
dimensionless parameter $\lambda$ controls the coupling
between the dilaton and the Maxwell fields.

\par
In the present paper we consider the static spherically symmetric
stringy charged black hole and calculated the energy distribution,
using the M{\o}ller energy-momentum complex. The result obtained
is an acceptable one, because the black hole under
consideration is charged, and it differs from the results obtained by
Virbhadra and Parikh \cite{VP93} and Xulu \cite{X98}. Both of them
provide the same energy, using Einstein and Tolman prescriptions,
respectively, in the space-time under consideration which is given
by $E = M$.
\par
Our result sustains two directions; the first, the viewpoint of
Lassner \cite{L96}  that the M{\o}ller energy-momentum complex is
a powerful concept of energy and momentum. The second, the results
obtained by Xulu \cite{grO0} and Virbhadra \cite{V99} that the
energy distribution in the sense of Einstein and M{\o}ller
disagree in general.

}
\end{document}